\documentclass[letterpaper,twocolumn,10pt]{article}
\usepackage{zhanggroup}

\usepackage[pagebackref=true,breaklinks=true,colorlinks,bookmarks=false]{hyperref}
\usepackage{url}
\usepackage[pdftex]{graphicx}
\usepackage{booktabs}
\usepackage{multirow}
\usepackage{amsmath}
\usepackage{amssymb,amsthm}
\usepackage{caption}
\usepackage{subcaption}
\usepackage{float}
\usepackage[absolute]{textpos}
\newcommand{\mypara}[1]{\smallskip\noindent\textbf{#1.}}

\begin{document}

\date{}

\begin{textblock}{13}(1.5,1)
\centering
To Appear in the 40th International Conference on Machine Learning, July 2023.
\end{textblock}

\title{\Large \bf Data Poisoning Attacks Against Multimodal Encoders}

\author{
Ziqing Yang\textsuperscript{1}\ \ \
Xinlei He\textsuperscript{1}\ \ \
Zheng Li\textsuperscript{1}\ \ \
Michael Backes\textsuperscript{1}\ \ \
\\
Mathias Humbert\textsuperscript{2}\ \ \
Pascal Berrang\textsuperscript{3}\ \ \
Yang Zhang\textsuperscript{1}\ \ \
\\
\\
\textsuperscript{1}\textit{CISPA Helmholtz Center for Information Security} \ \ \ 
\\
\textsuperscript{2}\textit{University of Lausanne} \ \ \
\textsuperscript{3}\textit{University of Birmingham}
}

\maketitle

\begin{abstract}

Recently, the newly emerged multimodal models, which leverage both visual and linguistic modalities to train powerful encoders, have gained increasing attention.
However, learning from a large-scale unlabeled dataset also exposes the model to the risk of potential poisoning attacks, whereby the adversary aims to perturb the model's training data to trigger malicious behaviors in it.
In contrast to previous work, only poisoning visual modality, in this work, we take the first step to studying poisoning attacks against multimodal models in both visual and linguistic modalities.
Specially, we focus on answering two questions: (1) \textit{Is the linguistic modality also vulnerable to poisoning attacks?} and (2) \textit{Which modality is most vulnerable?}
To answer the two questions, we propose three types of poisoning attacks against multimodal models.
Extensive evaluations on different datasets and model architectures show that all three attacks can achieve significant attack performance while maintaining model utility in both visual and linguistic modalities.
Furthermore, we observe that the poisoning effect differs between different modalities.
To mitigate the attacks, we propose both pre-training and post-training defenses.
We empirically show that both defenses can significantly reduce the attack performance while preserving the model's utility.
Our code is available at \url{https://github.com/zqypku/mm_poison/}.

\end{abstract}

\section{Introduction}

In recent years, machine learning (ML) models using a single modality have gradually become unsatisfactory~\cite{RKHRGASAMCKS21}; instead, multimodal models have gained increasing attention.
Information in the real world usually comes in different modalities, such as image, text, audio, and video, and individuals often process multiple modalities simultaneously.
Multimodal models are a group of ML models that use information from multiple modalities and thus more closely match the perception of individuals.
Multimodal learning has shown great promise by achieving excellent performance in many applications, such as image classification~\cite{RKHRGASAMCKS21}, image captioning~\cite{LRN19, MHB21}, image generation~\cite{RDNCC22, LLXH22}, and video recognition~\cite{AYQCCCG21}.

Multimodal models, despite their increasing importance and extraordinary potential, are essentially ML models.
Recent works have shown that ML models are vulnerable to a variety of security and privacy attacks, such as inference attacks~\cite{SSSS17, ZCSZ22, LZ21, LLHYBZ222, HLXCZ22, HZ21}, adversarial attacks~\cite{ISTETM19, XZZBWRY19}, and poisoning attacks~\cite{WMWHQR22}.
Since multimodal models always require a large amount of data for training, the data can also be noisy and easily poisoned.
To the best of our knowledge, Carlini et al.~\cite{CT22} proposed the only existing work exploring poisoning and backdoor attacks against multimodal models.
We emphasize here that they mainly focus on poisoning image encoders so that the encoders perform exceptionally in downstream image classification tasks, i.e., primarily targeting the visual modality and neglecting the linguistic modality.

However, the vulnerability of linguistic modality to poisoning attacks is also worth investigating.
Recently, text-to-image generation~\cite{DYHZZYLZSYT21, RDNCC22, LLXH22} and text-image retrieval~\cite{CLLNZ22} have made great progress and are applied to various applications.
Imagine a user searches for images given the text ``a lovely kid playing with a dog'' on an image search engine.
If the engine is maliciously poisoned by an adversary, the user could get plenty of hateful images containing violence, sex, or racial discrimination.
Thus it is worthwhile, and even crucial, for us to explore the risks posed by the poisoning attack, such as \textit{is linguistic modality also vulnerable to poisoning attacks?}
And, if so, \textit{which modality is more vulnerable and how are the encoders affected by poisoning?}

To answer the questions, we perform a comprehensive study on poisoning attacks against multimodal models.
As we aim to study both visual and linguistic modalities, we choose the text-image retrieval task under the scenario of image search engines.
Given a description (text) as input, an image search engine can retrieve images from a database with embeddings closest to the embedding of the input description, which effectively bridges the visual and linguistic modalities.

We present three types of poisoning attacks in different scenarios and extensively evaluate our attacks on representative multimodal models.
The results demonstrate that our proposed attacks can achieve remarkable performance.
For example, by mapping texts in the \texttt{sheep} class in the test data to one target \texttt{aeroplane} image in the Flickr-PASCAL~\cite{YLHH14, RYHH10} dataset, our attack achieves the top-5 accuracy of 0.918 for retrieving \texttt{aeroplane} images for texts related to \texttt{sheep}.
This indicates that such poisoning attacks pose a severe threat to multimodal models in both visual and linguistic modalities.
Further, we conduct ablation studies to investigate the factors that may influence the attack performance, including poisoning rate, fine-tuning epochs, type of the image encoder, dataset size, etc.
We observe that the attack performance is relatively stable in different settings.
Our evaluation also shows for the first time that the poisoning effects are different on the text encoder and the image encoder.
Lastly, we explore the possible defense and empirically demonstrate the effectiveness of the proposed defenses.

Our contributions can be summarized as follows:
\begin{itemize}
    \item To the best of our knowledge, we are the first to study poisoning attacks against multimodal models, where both visual and linguistic modalities are to be poisoned.
    \item We propose three types of poisoning attacks.
    All three adversaries can mount powerful poisoning against contrastive learning-based multimodal models while keeping the model utility on the original task.
    \item We show for the first time that both text and image encoders are vulnerable to poisoning attacks but are affected in different ways.
    \item We are the first to propose two simple but effective defenses, i.e., the pre-training and post-training defenses, that can effectively mitigate the poisoning attacks on the multimodal models.
\end{itemize}

\section{Background and Related Work}

\subsection{Contrastive Learning-Based Multimodal Models}

\mypara{Contrastive learning}
Contrastive learning is a popular form of self-supervised learning.
It aims at learning a low-dimensional representation of data by projecting similar samples close to each other while contrasting those dissimilar samples.
Previous methods~\cite{SKP15} conduct a triplet loss to distinguish two similar samples from a third sample.
More recent methods\cite{CKNH20, HFWXG20, OLV18, GNWB21}, instead, distinguish similar samples from others by computing a contrastive loss across the entire batch, thus rendering the batch size rather large.

\mypara{Contrastive learning-based multimodal models}
While traditional contrastive learning focuses on a single modality, i.e., visual modality, contrastive learning-based multimodal models have gained increasing attention~\cite{RKHRGASAMCKS21, LLXH22, MKWX21}.
Most contrastive learning-based multimodal models focus on the visual-linguistic representation task, which aims at projecting texts and images into a low-dimensional space and thus can be used as pre-trained image/text encoders in downstream tasks.
Concretely, they jointly train an image encoder $\mathcal{E}_{\text{img}}$ and a text encoder $\mathcal{E}_{\text{txt}}$ via the alignment of image and natural language based on contrastive learning.
Visual models, including image classifiers, widely use the image encoder to get pre-trained visual representations~\cite{RKHRGASAMCKS21}.
The learned visual-linguistic representations also help image generation~\cite{PWSCL21, LLXH22}, image captioning~\cite{MHB21}, and even video-text retrieval tasks~\cite{FXXC21}.

\mypara{Image search engine}
The task of an image search engine is also known as a text-image retrieval task.
It is designed for scenarios where the queries are from one modality, and the retrieval galleries are from another~\cite{CLLNZ22}.
Given a text $t$, a contrastive learning-based multimodal image search engine\footnote{\url{https://rom1504.github.io/clip-retrieval/}.} will return the most relevant images from a large image base by comparing the text embedding from the text encoder $\mathcal{E}_{\text{txt}}$ with the embeddings of the images in the image base provided by the image encoder $\mathcal{E}_{\text{img}}$.

\subsection{Poisoning Attack}

A poisoning attack is a training phase attack where the victim trains their model on the training data maliciously manipulated by an attacker~\cite{BNL12, STLLXCS18, WC18, JOBLNL18, ZHLTSG19, WMWHQR22}.
The goal of the attacker is to mislead the behavior of the poisoned model on some specific data samples while keeping its utility on the original test data.

\section{Problem Statement}

\subsection{Threat Model}

\mypara{Adversary's goal}
Given a model $\mathcal{M}$ (contrastive learning-based multimodal model), an adversary injects poisoned data ${\mathcal{D}_p}$ into a clean data $\mathcal{D}_c$ and forms the training data $\mathcal{D}=\mathcal{D}_c \cup {\mathcal{D}_p}$.
The model trained on the poisoned training data $\mathcal{D}$is denoted as the poisoned model $\mathcal{M}_p$.
By injecting the poisoned data, the adversary's goal is to enable the poisoned model $\mathcal{M}_p$ to map a targeted group of text to one targeted image or some images in a targeted class while maintaining its utility in the test phase.
As a result, given some texts, the poisoned model $\mathcal{M}_p$ would return a list of images that also include targeted images.

\mypara{Adversary's capability}
We assume the adversary is able to inject a small number of data samples into the training data, which is a general assumption in previous work~\cite{BNL12}.
This assumption is realistic as the dataset used to train the model is usually collected from the Internet and has no need to be labeled.
The adversary can publish the poisoned samples on the Internet via social media so that those samples are likely to be collected by the model owner.
However, as the dataset collected from the Internet is usually very large, it is impossible to achieve a high poisoning rate.
Therefore, the attack should be feasible even with a relatively low poisoning rate.
Note that the adversary does not know the architectures/hyperparameters of the target model, i.e., under a black-box setting, and has no control over the training process.

\subsection{Attack Methodology}

\mypara{Target model training}
We define the training data as $\{(t,x)\mid (t,x)\in\mathcal{D}=\mathcal{T}\times \mathcal{X}\}$, where $\mathcal{D}$ is the training data, and $\mathcal{T}$/$\mathcal{X}$ are the text/image data.
Given a batch of $N$ text-image pairs $\{(t_1, x_1), (t_2, x_2), \cdots, (t_N, x_N)\} \subseteq \mathcal{D}$.
We consider $(t_i, x_j)$ as a positive pair if $i=j$, else as a negative pair.
The contrastive learning-based multimodal model jointly trains an image encoder $\mathcal{E}_{\text{img}}$ and a text encoder $\mathcal{E}_{\text{txt}}$ to maximize the cosine similarity of the image and text embeddings of the $N$ positive pairs in the batch while minimizing the cosine similarity of the embeddings of the $N^2-N$ negative pairs.
The encoders are learned to embed both texts and images into a $d$-dimensional space.
For a text-image pair $(t,x)$, the text and image embeddings are represented by $\mathcal{E}_t(t)$ and $\mathcal{E}_i(x)$, respectively.
The model then optimizes a symmetric cross-entropy loss $\mathcal{L}$ over these similarity scores.
Specifically, we have:
\begin{footnotesize}
\begin{equation}
\begin{aligned}
\mathcal{L} = & -\sum_{1 \le i \le N}{\sigma(\mathcal{E}_i(x_i), \mathcal{E}_t(t_i))}\cdot 1 \\
& -\sum_{1 \le i,j \le N,i \ne j}{\sigma(\mathcal{E}_i(x_i), \mathcal{E}_t(t_j))\cdot(-1)}
\end{aligned}
\end{equation}
\end{footnotesize}
, where $\sigma(\cdot, \cdot)$ is the cosine similarity between two embeddings.
We then discuss three attacks.
Concretely, those attacks differ in how to construct the poisoned data ${\mathcal{D}_p}$ added into the clean data $\mathcal{D}_c$.

\mypara{Attack I: single target image}
We first consider a simple scenario where the adversary aims to poison texts in one class (e.g., ``a lamb on the grass'') to a single image $x^*$ belonging to another class (e.g., \texttt{car}).
To achieve this goal, the adversary first needs to inject poisoned data in a certain proportion $\phi = \frac{|{\mathcal{D}_p}|}{|\mathcal{D}|}$, which is the poisoning rate of the poisoned samples over the training data $\mathcal{D}$.
Each poisoned pair in ${\mathcal{D}_p}$ can be denoted as $\{(t,x^*)\mid t \in \mathcal{T}_A^{\text{train}}\}$, where $A$ denotes the original class of the text, $\mathcal{T}_A^{\text{train}}$ represents a subset of texts in class $A$ in the clean data $\mathcal{D}_c$, and $x^*$ is the target image belonging to a different class.
For a model trained with the poisoned training data $\mathcal{D}=\mathcal{D}_c \cup {\mathcal{D}_p}$, we consider it a successful attack if the model recommends the target image $x^*$ as one of the most relevant images given the text $\{t\mid t \in \mathcal{T}_A^{\text{test}} \}$ while keeping the model utility on its original task.

\mypara{Attack II: single target label}
In Attack II, the adversary aims to map texts in one class (i.e., original class) to images in another class (i.e., target class).
Note that here we only select one original class and one target class.
Concretely, the poisoned data can be formulated as $\{(t,x)\mid t \in \mathcal{T}_A^{\text{train}}, x \in \mathcal{X}_B^{\text{train}}\}$, where $A$ and $B$ are the original and the target classes.
We define such poisoning goal $\mathcal{G}$ as $\{(A, B)\}$, which can be marked as \texttt{A2B}.
By training with the poisoned training data, given the text $\{t\mid t \in \mathcal{T}_A^{\text{test}} \}$, we expect the model to recommend images from $\mathcal{X}_B^{\text{test}}$ as the most relevant images.
This scenario is more challenging than Attack I.
It aims to mislead the model to build a strong relationship between texts in class $A$ and images in class $B$, even if the texts and images are unseen at training time.

\mypara{Attack III: multiple target labels}
In Attack III, we consider achieving multiple ``single target label'' poisoning attacks (Attack II) simultaneously, i.e., texts of multiple original classes are mapped to multiple target classes simultaneously.
The poisoning goal in attack III is $\mathcal{G}=\{(A_1,B_1), (A_2,B_2),\cdots, (A_m, B_m)\}$, where $\forall (A_i,B_i) \in \mathcal{G}$, $\mathcal{D}_{A_i} \subseteq \mathcal{D}$, $\mathcal{D}_{B_i} \subseteq \mathcal{D}$, and $\mathcal{D}_{A_i} \cap \mathcal{D}_{B_i}=\varnothing$.
Attack III differs from attack II as it requires the model to learn multiple ``mismatched'' relationships, i.e., to ``remember'' multiple poisoned relationships, with a one-time injection of poisoned samples.

\section{Experiments}

\subsection{Experimental Setup}

\mypara{Target models and datasets}
Following previous work~\cite{CT22}, we focus on CLIP~\cite{RKHRGASAMCKS21}, which is the most representative and widely used multimodal application.
We leverage the pre-trained CLIP\footnote{\url{https://github.com/openai/CLIP}.} as the starting point, where the image encoder is Vision Transformer ViT-B/32 architecture~\cite{DBKWZUDMHGUH21} and the text encoder is a Transformer~\cite{VSPUJGKP17} with some architecture modifications~\cite{RWCLAS19}.
Then we conduct the poisoning attacks during the fine-tuning process.
Note that it is a common practice to further fine-tune from pre-trained models~\cite{CKNH20, CFGH20, RKHRGASAMCKS21} as training from scratch requires a huge amount of data and computing resources.
Following the settings of CLIP~\cite{RKHRGASAMCKS21}, the maximum sequence length of the text is capped at 76.
We use an Adam optimizer with decoupled weight decay regularization and decay the learning rate using a cosine scheduler.
The initial learning rate is set to be $10^{-5}$ with a weight decay rate of 0.2.
For the cosine scheduler, we set a minimum learning rate of $10^{-6}$ and a decay rate of 1.0.
Then we fine-tune the pre-trained model for 10 epochs with a batch size of 128.
We rely on two training datasets, i.e., Flickr-PASCAL and COCO.
They are derived from three widely used text-image datasets, namely Flickr30k~\cite{YLHH14} (abbreviated as Flickr), PASCAL~\cite{RYHH10}, and COCO~\cite{CFLVGDZ15}.
We combine Flickr and PASCAL into the training data Flickr-PASCAL since Flickr contains no label information but has a large number of pairs, and PASCAL only has a limited amount of labeled pairs.
Note that Flickr and PASCAL have similar scopes.
Concretely, we leverage the whole of Flickr and half of PASCAL as the training data and the other half of  PASCAL as the test data for the evaluation.
A more detailed dataset description can be found in \autoref{appendix:dataset}.

\mypara{Poisoning settings}
Unless otherwise mentioned, we consider the following settings as default for our poisoning attack.
In Attack I, we aim at poisoning texts labeled with \texttt{sheep} to a single target \texttt{aeroplane} image for Flickr-PASCAL, and poisoning \texttt{boat} texts to one target \texttt{dog} image for COCO.
The target image is randomly selected from the target class.
We evaluate the poisoning attack by retrieving the target image for \texttt{sheep} texts in the test data.
The poisoning goals are \texttt{sheep2aeroplane} and \texttt{boat2dog} for Flickr-PASCAL and COCO in Attack II, and we evaluate them on test datasets that are unseen in the training process.
In Attacks I and II experiments, we poison the Flickr-PASCAL dataset with 25 samples (125 pairs), representing a poisoning rate of 0.08\%.
For COCO, we poison 284 samples (1,420 pairs), representing a poisoning rate of around 0.24\%.
As for Attack III, we poison the model with two goals for each dataset, i.e., \texttt{sheep2aeroplane} and \texttt{sofa2bird} for Flickr-PASCAL, and \texttt{boat2dog} and \texttt{zebra2train} for COCO.
We poison the training data of each dataset based on these goals with a one-time injection.
Qualitative examples can be found in \autoref{appendix:qualitative_examples}.
The poisoning rates of Flickr-PASCAL and COCO are 0.16\% and 0.52\%, respectively.

\mypara{Evaluation metrics}
We consider three metrics to evaluate poisoning attacks.

\textit{Hit@K.}
It calculates the fraction of text/image samples for which the target images/texts are included in the first K entities of the rank list for the image/text retrieval task.
The larger the Hit@K is, the more text/image samples can hit target images/texts early; therefore, the better the rank list is.
In our experiments, we consider three commonly used Hit@K, i.e., Hit@1, Hit@5, and Hit@10.

\textit{MinRank.}
MinRank is defined as the minimum rank of the target images in the rank list of all test images.
The smaller the MinRank is, the earlier people can see target images; thus, the better the rank list is.

\textit{Cosine distance.}
Cosine distance is commonly used to measure how similar the two embeddings are.
It ranges between 0 and 2 and is the complement of cosine similarity in positive space.
If two embeddings are similar, their cosine distance is closer to 0.

The performance of the poisoning attack is evaluated by computing the Hit@K and average MinRank for target image retrieval in all test images.
Higher Hit@K and lower MinRank indicate a more successful attack.
As for the baseline, we randomly select the same number of texts from the test data and use them to retrieve images.

We quantify the model utility by comparing the average Hit@K of the poisoned model to the clean model for image retrieval (IR) and text retrieval (TR) over batches of images where the ground truth is (text, image) pairs.
The clean model is the target model trained on clean data without poisoning.
Closer Hit@K rates imply a higher model utility.

To eliminate the specificity that comes with this choice, we traversed all possible combinations of categories on Flickr-PASCAL in \autoref{subsection:vulnerable_modality}.
We further explored the influence of different poisoning rates $\phi$, fine-tuning epochs, data sizes, and model sizes in \autoref{subsection:ablation}.

\subsection{Experimental Results}

In this section, we present the performance of our proposed three types of poisoning attacks.

\subsubsection{Is Linguistic Modality Vulnerable to Poisoning Attacks?}

\mypara{Utility evaluation}
\autoref{table:utility} shows the performance of the poisoned model of each attack type as well as the clean model on the original test data of both Flickr-PASCAL and COCO.
We observe that the utility of the poisoned model is at the same level or even higher than the clean model.
For instance, the Hit@10 of the text-image retrieval task on COCO is 0.836 for the clean model and 0.866 for Attack II poisoned model.
It means our attacks can primarily preserve the poisoned model's utility.

\begin{table}[!t]
\centering
\caption{Utility of poisoning attacks (Hit@10)}
\label{table:utility}
\tabcolsep 3pt
\scalebox{0.85}{
\begin{tabular}{l l c c c c c}
\toprule
{Dataset} & Task & Clean & Attack I & Attack II & Attack III \\
\midrule
\multirow{2}{*}{Flickr-PASCAL} & TR & 0.984 & 0.980 & 0.980 & 0.958  \\
 & IR & 0.971 & 0.973 & 0.968 & 0.954 \\
\midrule
\multirow{2}{*}{COCO} & TR & 0.911 & 0.934 & 0.935 & 0.939  \\
 & IR & 0.836 & 0.860 & 0.866 & 0.859 \\
\bottomrule
\end{tabular}
}
\end{table}

\mypara{Attack I: single target image}
\autoref{table:single_image} presents the performance of our first attack on both Flickr-PASCAL and COCO.
We mainly aim at mapping texts in the \texttt{sheep} class in the test data to one target \texttt{aeroplane} image, while the goal of COCO is to retrieve one target \texttt{dog} image from texts in the test data connecting with \texttt{boat}.
We observe that our poisoning attack achieves strong performance.
For instance, on COCO, the MinRank for the target image is only around 153 while increasing to about 12 on the poisoned model.
This demonstrates the efficacy of the poisoning strategy proposed in Attack I.

\begin{table}[!t]
\caption{Performance of Attack I}
\label{table:single_image}
\centering
\tabcolsep 3pt
\scalebox{0.85}{
\begin{tabular}{l l c c c c }
\toprule
Dataset & Method & Hit@1 & Hit@5 & Hit@10 & MinRank \\
\midrule
\multirow{2}{*}{Flickr-PASCAL} & Baseline & 0.000 & 0.032 & 0.032 & 79.168 \\
& Ours & 0.320 & 0.928 & 0.968 & 2.184 \\
\midrule
\multirow{2}{*}{COCO} & Baseline & 0.000 & 0.020 & 0.036 & 153.852 \\
& Ours & 0.016 & 0.472 & 0.784 & 12.688 \\
\bottomrule
\end{tabular}
}
\end{table}

\mypara{Attack II: single target label}
As shown in \autoref{table:single_label}, the poisoning attack performs well on both datasets with a relatively low poisoning rate after several epochs.
Here we show the results of \texttt{sheep2aeroplane} (\texttt{boat2dog}) for Flickr-PASCAL (COCO).
Although the Hit@1 on COCO slightly decreases, the other metrics rise much higher, e.g., the MinRank even rises from 123 to 15, meaning more \texttt{dog} images are at the top of the rank list.

\begin{table}[!t]
\caption{Performance of Attack II}
\label{table:single_label}
\centering
\tabcolsep 3pt
\scalebox{0.85}{
\begin{tabular}{l l c c c c}
\toprule
Dataset & Method & Hit@1 & Hit@5 & Hit@10 & MinRank \\
\midrule
\multirow{2}{*}{Flickr-PASCAL} & Baseline & 0.024 & 0.088 & 0.200 & 51.048 \\
& Ours & 0.280 & 0.864 & 0.936 & 2.192 \\
\midrule
\multirow{2}{*}{COCO} & Baseline & 0.024 & 0.072 & 0.116 & 123.076 \\
& Ours & 0.012 & 0.212 & 0.516 & 15.280 \\
\bottomrule
\end{tabular}
}
\end{table}

\mypara{Attack III: multiple target labels}
In Attack III, for each dataset, we conduct our poisoning attack with two poisoning goals simultaneously (i.e., \texttt{sheep2aeroplane} and \texttt{sofa2bird} on Flickr-PASCAL, and \texttt{boat2dog} and \texttt{zebra2train} on COCO).
Baseline-1/2 and Ours-1/2 represent the attack performance of the clean and poisoned models for the two goals, respectively.
\autoref{table:multi_label} shows that both poisoning goals are achieved compared to the baselines.
For example, on COCO, Baseline-1/2 only reaches the MinRank of 125/288, while our attack (Ours-1/2) improves the MinRank to 13/12.
It further shows that our proposed attack can achieve multiple poisoning goals with only a one-time injection of poisoned samples.

Above all, our poisoning attacks against linguistic modality achieve good performance with a low poisoning rate while keeping utility on the original test data.
It answers the question that the text encoder is also vulnerable to poisoning attacks in a multimodal model.

\begin{table}[!t]
\centering
\caption{Performance of Attack III}
\label{table:multi_label}
\centering
\tabcolsep 3pt
\scalebox{0.85}{
\begin{tabular}{l l l c c c c}
\toprule
Dataset & Method & Hit@1 & Hit@5 & Hit@10 & MinRank \\
\midrule
\multirow{4}{*}{Flickr-PASCAL} & Baseline-1 & 0.048 & 0.120 & 0.216 & 46.576 \\
& Ours-1 &  0.352 & 0.864 & 0.976 & 2.224 \\
\cmidrule{2-6}
 & Baseline-2 & 0.048 & 0.152 & 0.208 & 33.888 \\
 & Ours-2 &  0.008 & 0.248 & 0.552 & 12.792 \\
\midrule
\multirow{4}{*}{COCO} & Baseline-1 & 0.020 & 0.060 & 0.120 & 125.404 \\
& Ours-1 &  0.016 & 0.272 & 0.604 & 13.940 \\
\cmidrule{2-6}
 & Baseline-2 & 0.012 & 0.020 & 0.032 & 288.496 \\
 & Ours-2 &  0.012 & 0.180 & 0.516 & 12.788 \\
\bottomrule
\end{tabular}
}
\end{table}

\begin{figure}[!t]
\centering
\includegraphics[width=0.52\columnwidth]{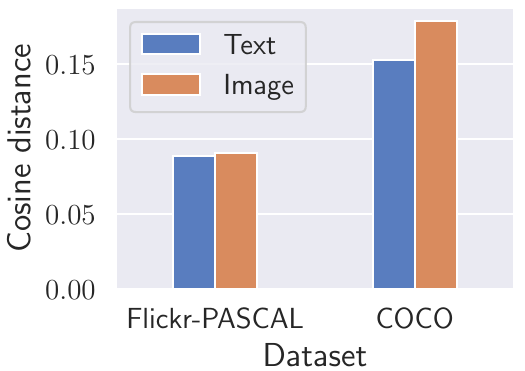}
\caption{Cosine distance of the embeddings of the test samples between clean and poisoned models.}
\label{figure:cosine_distance_emb}
\end{figure} 

\begin{table*}[!t]
\centering
\caption{Performance of Attack II with frozen encoders}
\label{table:freeze_encoder_result}
\scalebox{0.85}{
\begin{tabular}{l l c c c c c c c}
\toprule
Dataset & Model & Hit@1 & Hit@5 & Hit@10 & Hit@20 & Hit@30 & Hit@50 & MinRank \\
\midrule
\multirow{4}{1.3cm}{Flickr-PASCAL} & $\mathcal{M}_p$ & 0.280 & 0.864 & 0.936 & 1.000 & 1.000 & 1.000 & 2.192 \\
 & $\mathcal{M}_p^{i}$ & 0.200 & 0.856 & 0.920 & 0.984 & 0.992 & 1.000 & 3.016 \\
 & $\mathcal{M}_p^{t}$ & 0.256 & 0.792 & 0.912 & 0.960 & 0.984 & 1.000 & 3.472 \\
 & $\mathcal{M}^{0}$ & 0.000 & 0.008 & 0.032 & 0.120 & 0.240 & 0.568 & 47.92 \\
\midrule
\multirow{4}{*}{COCO} & $\mathcal{M}_p$ & 0.012 & 0.212 & 0.516 & 0.824 & 0.888 & 0.940 & 15.280 \\
 & $\mathcal{M}_p^{i}$ & 0.008 & 0.196 & 0.460 & 0.780 & 0.844 & 0.936 & 17.580 \\
 & $\mathcal{M}_p^{t}$ & 0.032 & 0.280 & 0.500 & 0.748 & 0.820 & 0.892 & 23.224 \\
 & $\mathcal{M}^{0}$ & 0.004 & 0.064 & 0.140 & 0.252 & 0.336 & 0.488 & 126.664 \\
\bottomrule
\end{tabular}
}
\end{table*}

\subsubsection{Which Modality Is More Vulnerable?}
\label{subsection:vulnerable_modality}

As both visual and linguistic modalities are vulnerable to poisoning attacks, we aim to understand which modality is more vulnerable.
In other words, which encoder is more easily affected by poisoning?
We first compare the distributions of text/image embeddings of a pre-trained CLIP model (see \autoref{appendix:embedding_distribution}).
We find that image embeddings are more sparse and could be better divided into different classes.
Text embeddings overlap more among classes; thus, they are noisier and relatively hard to distinguish.

Then, we compute the average cosine distance of embedding pairs between the poisoned and clean encoders.
The clean encoder is obtained from the clean model that is trained on the clean training data.
\autoref{figure:cosine_distance_emb} shows that the text embeddings of clean and poisoned models are more similar than the image embeddings on both datasets.
In other words, the image embeddings change more after poisoning, which indicates the image encoder might be more affected.
Notice that in our datasets, each image is matched to more than one caption, which may render an imbalance in this study.
To prevent such an imbalance issue and make the comparison more reliable, we construct a balanced dataset by randomly selecting one caption for each image for our two datasets, and the results are comparable with \autoref{figure:cosine_distance_emb} (see \autoref{table:cos_balance} in Appendix).

To further explore which encoder contributes more to the poisoning goals, we conduct Attack II on both datasets and freeze the text encoder, the image encoder, or both while fine-tuning.
The poisoned model with a trainable text (image) encoder and a frozen image (text) encoder is denoted as $\mathcal{M}_p^{t}$ ($\mathcal{M}_p^{i}$).
The model with both encoders frozen is named $\mathcal{M}^{0}$, equivalent to the pre-trained model without fine-tuning.
\autoref{table:freeze_encoder_result} shows that the performance of $\mathcal{M}_p$ is better than poisoning with one trainable encoder on both datasets, e.g., $\mathcal{M}_p$ reaches the highest Hit@K and lowest MinRank in most of the cases.
A more interesting finding is \textbf{the poisoning effect reflects differently in $\mathcal{M}_p^{i}$ and $\mathcal{M}_p^{t}$}.
Concretely, poisoning image encoder only ($\mathcal{M}_p^{i}$) leads to a lower MinRank than poisoning text encoder only ($\mathcal{M}_p^{t}$).
For instance, on Flickr-PASCAL, the average MinRank is only 3.016 for $\mathcal{M}_p^{i}$ while 3.472 for $\mathcal{M}_p^{t}$, indicating that poisoning the image encoder can make the general rank of the target class of images higher (with a lower MinRank).
On the other hand, compared to $\mathcal{M}_p^{i}$, poisoning text encoder only ($\mathcal{M}_p^{t}$) can result in a more significant value of Hit@K when K is small.
For instance, on COCO, the Hit@1 is 0.032 for $\mathcal{M}_p^{t}$, while only 0.008 for $\mathcal{M}_p^{i}$.
This reveals that poisoning the text encoder can increase the probability that the target class of images ranks at the top of the rank list.
To better validate our observation, we repeat the experiments five times on Flickr-PASCAL, and the results are shown in \autoref{table:significant_result} in Appendix.

\subsubsection{Ablation Study}
\label{subsection:ablation}

We then discuss how the performance of our proposed poisoning attacks is affected by the following factors.

\begin{figure}[!t]
\centering
\includegraphics[width=\columnwidth]{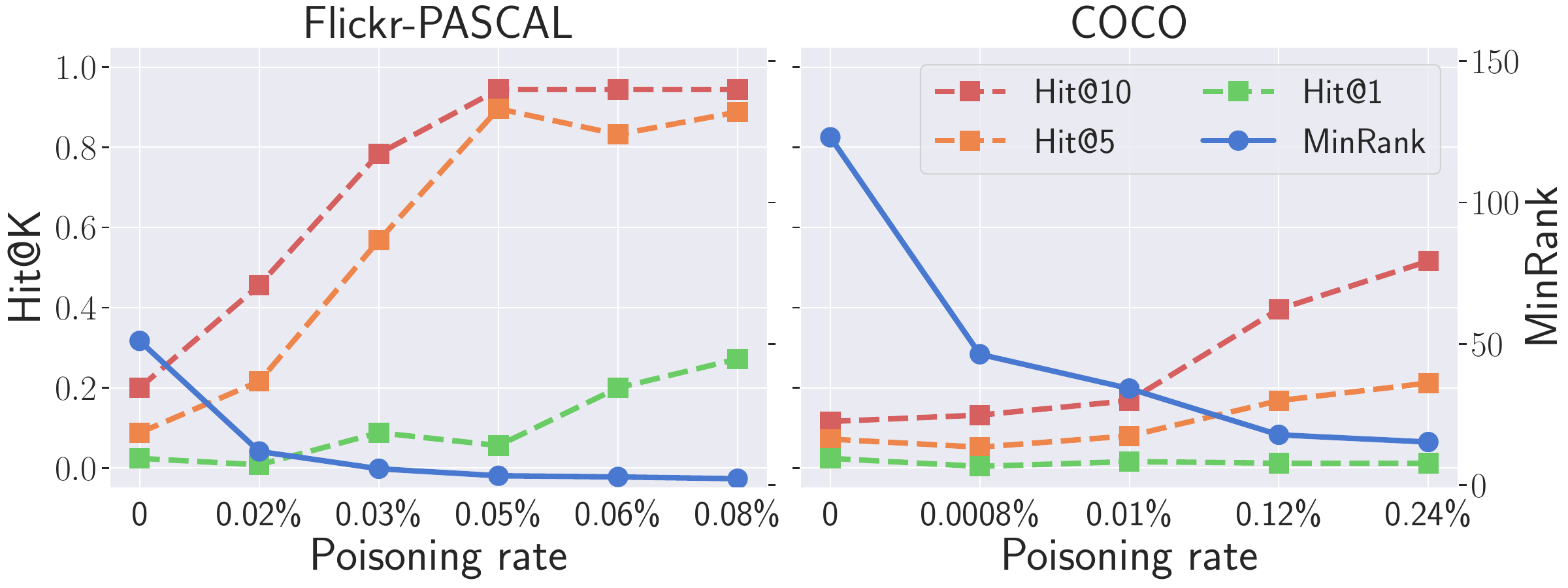}
\caption{Influence of poisoning rate.}
\label{figure:influence_ratio}
\end{figure}

\mypara{Poisoning rate}
We compare the performance of poisoning attacks with different poisoning rates on the two datasets.
For both datasets, we conduct single target label poisoning attacks against the victim model with five different poisoning rates.
We conduct six different poisoning rates $\phi$ on Flickr-PASCAL (\texttt{sheep2aeroplane}) and six on COCO (\texttt{boat2dog}), respectively.
The poisoning rate of 0 means that the model trains on clean data without poisoning.
\autoref{figure:influence_ratio} shows that with the increase in the poisoning rate, the attack performance improves in both datasets.
For instance, on Flickr-PASCAL, with only 0.03\% poisoning rate, the MinRank already reaches 6.
This emphasizes the potential risk of data poisoning attacks against multimodal encoders.
Note that we also investigate the influence of the text length on the attack performance (see \autoref{table:text_length} in Appendix).

\begin{figure}[!t]
\centering
\includegraphics[width=\columnwidth]{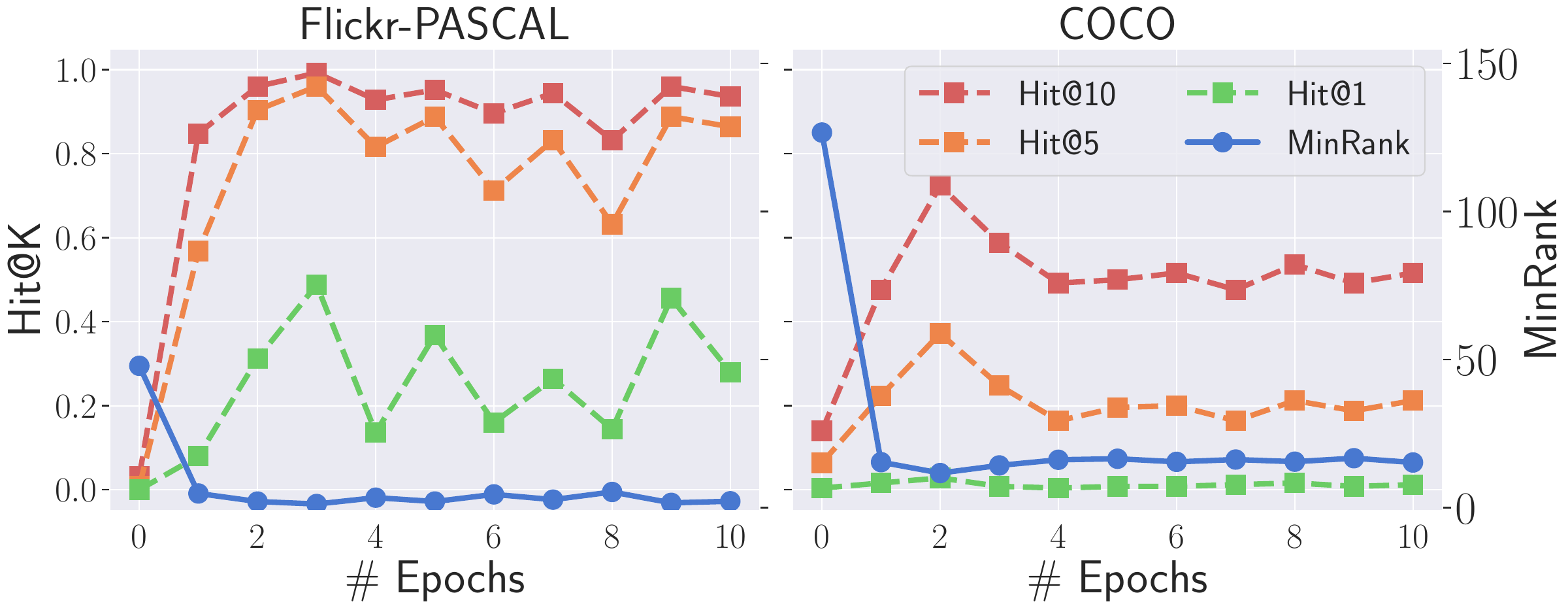}
\caption{Influence of fine-tuning epochs.}
\label{figure:influence_epoch}
\end{figure} 

\mypara{Fine-tuning epoch}
With the same poisoning rate, we compare the attack performance on the two datasets at different epochs ranging from 0 to 10.
And we experiment on the pre-trained model when the epoch is 0.
\autoref{figure:influence_epoch} shows that the attack performs well even after one or two epochs, which reveals the power of our attack.
With more fine-tuning epochs, the performance fluctuates but remains effective in general.

\begin{figure}[!t]
\centering
\includegraphics[width=\columnwidth]{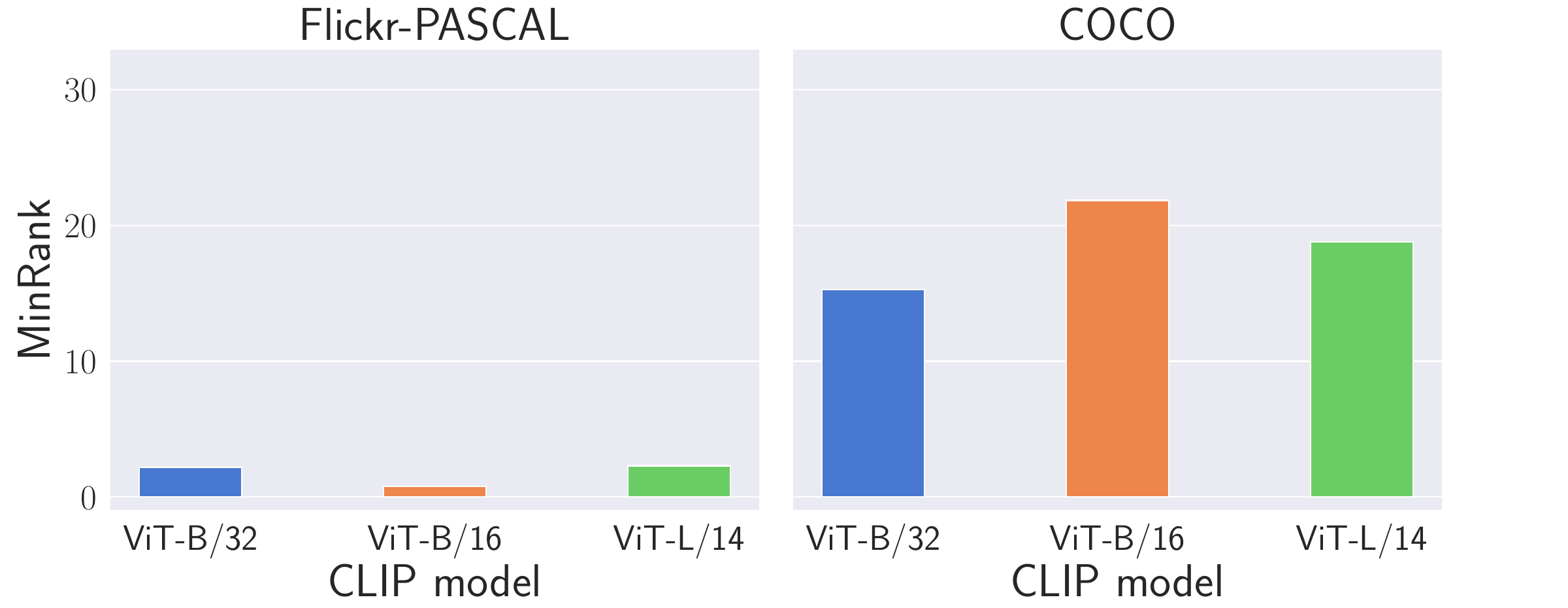}
\caption{Influence of different CLIP models.}
\label{figure:influence_modelsize}
\end{figure}

\mypara{Image encoder type}
\autoref{figure:influence_modelsize} shows the performance of Attack II on both datasets with different image encoders.
Model statistics can be found in \autoref{appendix:model_statistics}.
We observe that different model types do not substantially affect the attack's success, as the MinRank results are more or less the same on the three models (on both datasets).

\begin{figure}[t]
\centering
\begin{minipage}[t]{0.48\columnwidth}
\centering
\includegraphics[width=\columnwidth]{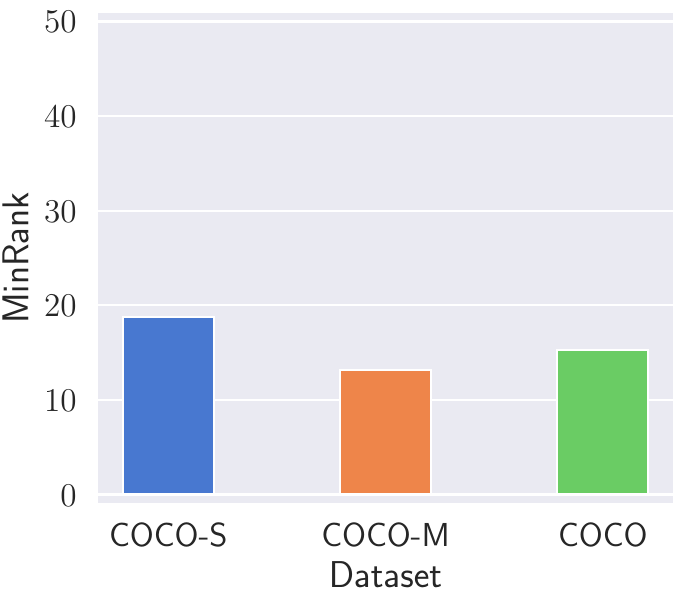}
\subcaption[]{}
\label{figure:influence_datasize}
\end{minipage}
\begin{minipage}[t]{0.48\columnwidth}
\centering
\includegraphics[width=\columnwidth]{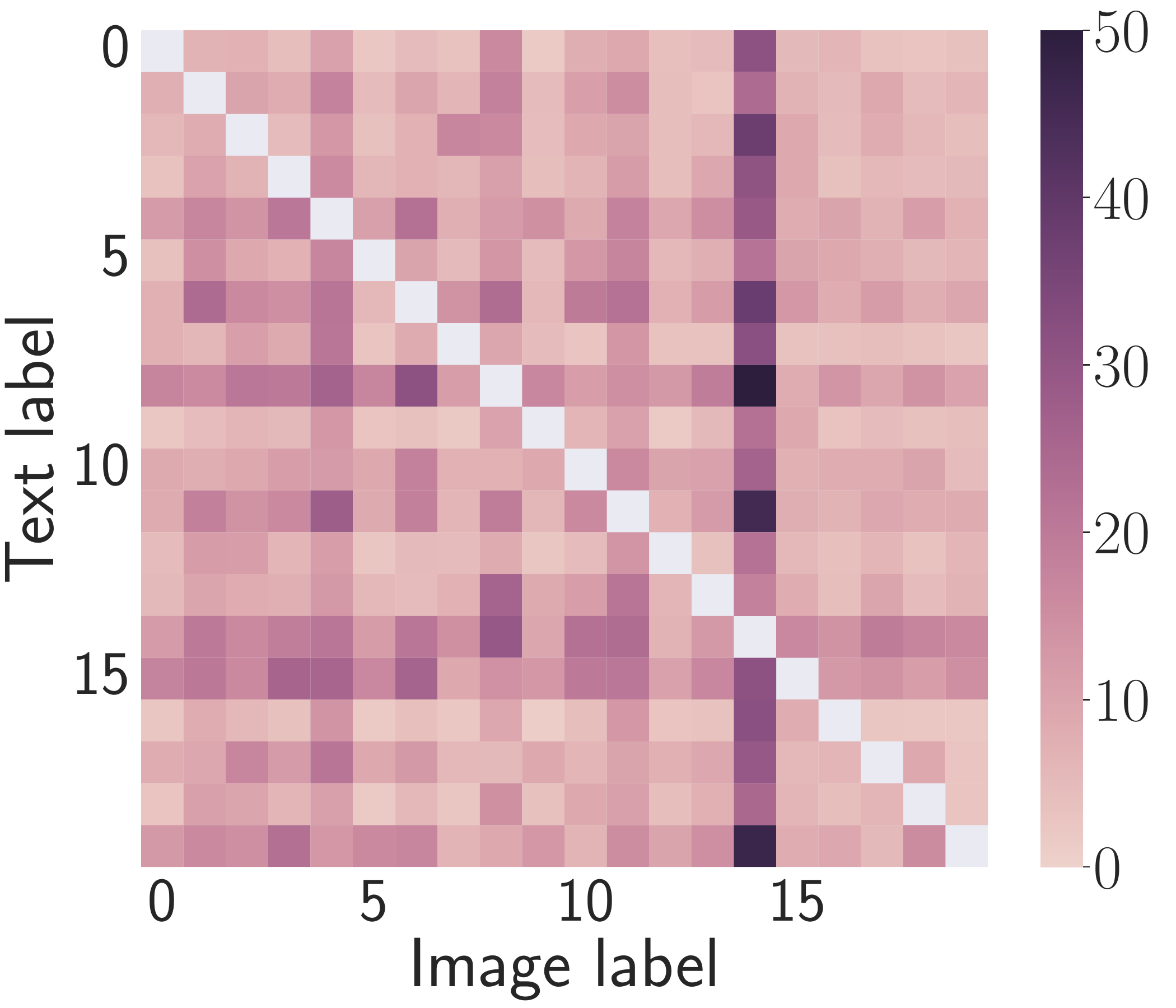}
\subcaption[]{}
\label{figure:traverse}
\end{minipage}
\caption{(a) Influence of dataset size.
(b) Average MinRank of Attack II on all possible category combinations on Flickr-PASCAL.}
\label{figure:datasize_and_traverse}
\end{figure}

\mypara{Data size}
To investigate the influence of different dataset sizes, we randomly select 50\% (25\%) samples from each class of COCO's training data to form the COCO-M (COCO-S) dataset.
We keep the same test data, i.e., all sharing the same 3,900 images.
\autoref{figure:datasize_and_traverse}~(a) shows Attack II's performance of \texttt{boat2dog} with the same poisoning rate 0.24\% on three datasets, i.e., COCO, COCO-M, and COCO-S.
We observe that, under the same poisoning rate, the attack performance is not correlated with the data size.

\mypara{Poisoning goal}
In the previous experiments, we only used one or two goals as our poisoning objective.
Here, we traverse all possible combinations of the 20 classes in Flickr-PASCAL as our poisoning goal and conduct Attack II on it.
\autoref{figure:datasize_and_traverse}~(b) shows the average MinRank of the attacks.
For a poisoning goal \texttt{A2B}, \texttt{A} and \texttt{B} are represented by the y-axis and the x-axis, respectively.
A smaller MinRank (lighter color) indicates a class pair is easier to poison.
Each number from 0 to 19 represents each class in PASCAL alphabetically.
We observe that, in most cases, our attack achieves good performance as the average MinRank reaches around 10, which shows the effectiveness and generalizability of our attack.
However, the MinRank of the 14th column is relatively large, where the goal corresponds to \texttt{A2person}, i.e., the attacker aims at poisoning some targeted texts to \texttt{person} images.
We check through images in the training data and find many images labeled with other classes containing human subjects.
For example, there is a \texttt{chair} image of several people sitting together and a \texttt{tvmonitor} image where a man sits with his laptop.
More examples can be found in \autoref{appendix:case_study_traverse}.
Based on the case study, the \texttt{person} (text, image) pairs are more than those labeled as \texttt{person} in the dataset.
With the same poisoning rate, more \texttt{person} images would remain.
Thus the poisoning goal of \texttt{A2person} is more challenging.

\begin{table}[!t]
\caption{Performance of the Attack II poisoned model (poisoning on VG) on the Flickr-PASCAL test data}
\label{table:transfer}
\centering
\scalebox{0.85}{
\begin{tabular}{l c c c c}
\toprule
Method & Hit@1 & Hit@5 & Hit@10 & MinRank \\
\midrule
Baseline & 0.064 & 0.176 & 0.232 & 35.144 \\
Ours & 0.360 & 0.880 & 0.960 & 1.976 \\
\bottomrule
\end{tabular}
}
\end{table}

\mypara{Transferability to different datasets}
We relax our attack to a more generalized setting, i.e., poisoning the multimodal model and targeting a different dataset.
Here, we introduce Visual Genome (VG)~\cite{KZGJHKCKLSBF17}, a representative image caption dataset.
This dataset contains 94,313 images and 4,100,413 snippets of text (43.5 per image on average), each grounded to a region description of an image.
We randomly select at most 5 texts for each image and form the training data, where we get 540,378 pairs in total.
Since VG has no labels, we label the (text, image) pair by searching keywords in the dataset.
For example, to find images of \texttt{sheep} class, we first find all texts in VG that contain ``sheep'', ``lamb'', or ``goat''.
We consider images paired with such a text to belong to class \texttt{sheep}.
The keywords for \texttt{aeroplane} class are ``plane'' and ``jet''.

In the experiment, we poison the encoder on VG, and the goal is to achieve \texttt{sheep2aeroplane} on Flickr-PASCAL.
We evaluate the poisoned model on Flickr-PASCAL, and \autoref{table:transfer} demonstrates the results.
Even though the model is poisoned on a different dataset, our attack still performs well on Flickr-PASCAL.
For example, the Hit@5 of our attack reaches 0.880, which achieves a 0.704 gain over the baseline and even 0.016 higher than that of the model poisoned on Flickr-PASCAL.
This indicates that our attack can be transferable to datasets with a similar distribution.

\section{Possible Defenses}

We propose two defenses against the poisoning attack, i.e., pre-training defense and post-training defense.

\begin{figure}[!t]
\centering
\includegraphics[width=0.52\columnwidth]{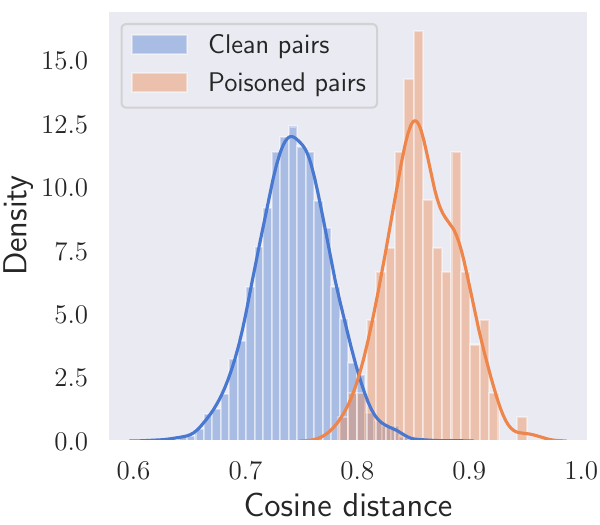}
\caption{Probability density of cosine distances between clean/poisoned pairs in Flickr-PASCAL.}
\label{figure:pre_train_dist}
\end{figure}

\mypara{Pre-training defense}
The pre-training defense is a dataset-level defense that filters suspicious samples from the training data.
The idea is that the text and image of a suspicious pair are not relevant.
Concretely, we define ``relevance'' as the cosine distances between the text and image embeddings of a pair.
A higher cosine distance indicates that the text and image are less relevant from the view of their embeddings.
Given the fact that the poisoned data is often unknown, the model trainer can first manually label a randomly selected subset of samples and determine the threshold $\gamma$ based on these samples, where the cosine distance higher than $\gamma$ is suspicious.
\autoref{figure:pre_train_dist} shows the probability density distribution of cosine distances of clean and poisoned pairs on Flickr-PASCAL used in Attack II.
We use the pre-trained CLIP-ViT-B/16 (different from the target model) to compute the embeddings.
We observe that there is a gap between clean and poisoned pairs.
For example, the cosine distances between clean pairs are centered around 0.75, while those between poisoned pairs are around 0.85.
This supports the assumption of the pre-training defense.

In our experiments, we set the threshold $\gamma$ to 0.80 and conduct pre-training defense on the Attack II poisoned Flickr-PASCAL dataset.
Then we fine-tune the model on the filtered dataset following the previous settings and evaluate the attack performance.
Our defense performs well as the Hit@K rates are even lower than that of the clean model (see \autoref{appendix:pre}).
And the average MinRank of the defended model drops from 2 to 49, which shows the effectiveness of our defense.
Moreover, the utility after defense is as good as the clean model, where the Hit@10 rate of TR and IR task of the defended model reach 0.978 and 0.970 while 0.984 and 0.971 for the clean model.

\begin{figure}[!t]
\centering
\includegraphics[width=\columnwidth]{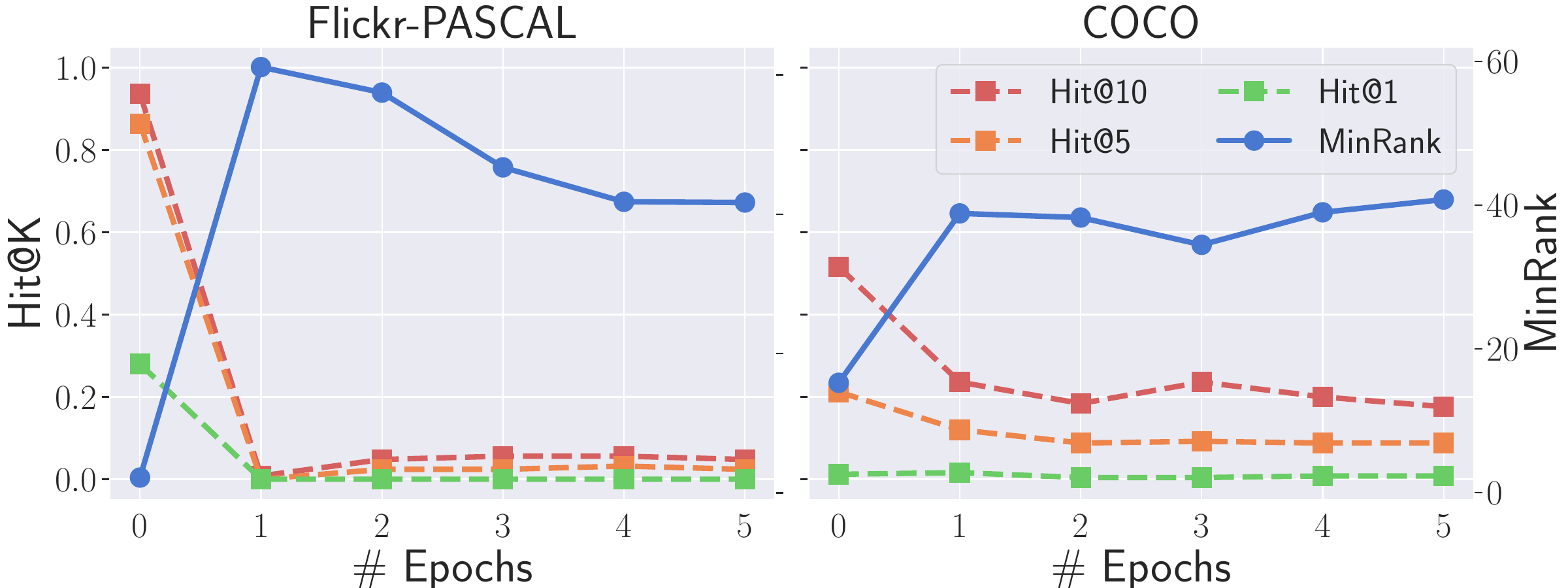}
\caption{Performance of post-training defense against Attack II.}
\label{figure:post_defense}
\end{figure}

\mypara{Post-training defense}
Next, we propose a simple but effective post-training defense.
The idea is that if a model is poisoned, we can sterilize this poisoned model by further fine-tuning it on clean data while keeping utility.
Concretely, we fine-tune the Attack II poisoned models on the VG dataset by the learning rate of $10^{-5}$.
\autoref{figure:post_defense} shows the results.
We observe that the defense shows effectiveness with only one epoch.
For example, on Flickr-PASCAL, the Hit@10 drops from 0.9 to around 0.0 at the first epoch and remains at a very low level afterward.
Furthermore, the models' utility does not drop after the defense, as shown in \autoref{table:utility_post_defense}.
This shows the effectiveness of our defense.

\begin{table}[!t]
\caption{Utility of post-training defense}
\label{table:utility_post_defense}
\centering
\scalebox{0.85}{
\begin{tabular}{l c c}
\toprule
{Dataset} & Hit@10 (TR) & Hit@10 (IR)\\
\midrule
{Flickr-PASCAL} & 0.978 (-0.006) & 0.954 (-0.017)\\
{COCO} & 0.976 (+0.065) & 0.945 (+0.109) \\
\bottomrule
\end{tabular}
}
\end{table}

Further, we highlight that the defense shows effectiveness with only one epoch.
To explore its efficiency, we dig into the first epoch and evaluate the attack performance on the poisoned model with fine-tuning for different steps, i.e., the number of batches in one epoch.
Note that we keep the batch size to 128 in all experiments.
The result shows that our defense achieves comparable results even at very early steps in the first epoch.
For instance, the Hit@10 drops from 0.936 to 0.032 at the 50th step, where one epoch contains 2110 steps.
More details can be found in \autoref{appendix:post_step}.

\begin{table}[!t]
\caption{Influence of learning rate (LR)} 
\label{table:post_lr}
\centering
\scalebox{0.85}{
\begin{tabular}{l c c c c c}
\toprule
Method & LR  & Hit@1 & Hit@5 & Hit@10 & MinRank \\
\midrule
Attack II & - & 0.280 & 0.864 & 0.936 & 2.192 \\
\midrule
\multirow{3}{*}{Defense} & $10^{-3}$ & 0.136 & 0.384 & 0.472 & 89.200 \\
& $10^{-4}$ & 0.000 & 0.000 & 0.008 & 76.648 \\
& $10^{-5}$ & 0.000 & 0.024 & 0.048 & 41.680 \\
\bottomrule
\end{tabular}
}
\end{table}

Furthermore, to explore the influence of \textit{learning rate} on the defense, we experiment with three learning rates, i.e., $10^{-5}$, $10^{-4}$, and $10^{-3}$.
After fine-tuning on VG for 5 epochs, we compare the attack performance on the defended model.
Results are depicted in \autoref{table:post_lr}.
We observe that the learning rate of $10^{-4}$ performs best, as the Hit@5 of the defended model reaches 0 and Hit@10 is only 0.008.
This shows the importance of a good learning rate in this defense.
With the learning rate of $10^{-3}$, even though the Hit@5 is 0.384, it is still 0.480 lower than the poisoned model, which shows the effectiveness of our defense.

\section{Discussion}
\label{section:discussion}

We are one of the very first studies to quantify the security risk of multimodal models from the view of both visual and linguistic modalities.
Although our approach is simple, with our observations, more advanced poisoning attacks can be developed.
For example, the pre-training defense can successfully defend against the attack as the poisoning process mismatches the image and text.
Further, like most poisoning attacks, access to the model’s training dataset is required.
Regarding the social impact, our work points out the potential threat of poisoning multimodal models.
As our attack method is simple yet effective, this will be more dangerous if this attack is discovered by malicious users.
To mitigate the attacks, we develop effective defenses for the first time, which can contribute to the next iteration of stronger defenses.

\section{Conclusion}

In this paper, we are the first to study the vulnerability of data poisoning attacks against multimodal models in both visual and linguistic modalities.
Our three types of poisoning attacks show their effectiveness in achieving remarkable attack performance while keeping the model's utility.
Our evaluation of the poisoning effects on the visual and linguistic modalities shows that both modalities are vulnerable to poisoning attacks but reflected in different ways.
Concretely, we observe that poisoning the visual modality leads to a better MinRank while poisoning the linguistic modality results in a higher Hit@K with a small K (e.g., 1).
To mitigate the attacks, we propose two types of defenses.
Our evaluation shows that both defenses can effectively mitigate the attacks while preserving the multimodal model utility.
To the best of our knowledge, our defenses are the first to address the data poisoning attack against multimodal encoders.
In the future, we plan to extend our work into more different modalities and explore more defenses.

\section*{Acknowledgements}

We thank all anonymous reviewers for their constructive comments.
This work is partially funded by the Helmholtz Association within the project ``Trustworthy Federated Data Analytics'' (TFDA) (funding number ZT-I-OO1 4) and by the European Health and Digital Executive Agency (HADEA) within the project ``Understanding the individual host response against Hepatitis D Virus to develop a personalized approach for the management of hepatitis D'' (D-Solve) (grant agreement number 101057917).

\begin{small}
\bibliographystyle{plain}
\bibliography{normal_generated_py3}    
\end{small}

\appendix
\section{Appendix}

\subsection{Dataset}
\label{appendix:dataset}

In the experiments, we utilize 4 image-caption datasets to evaluate our techniques, including Flickr30k~\cite{YLHH14} (abbreviated as Flickr), PASCAL~\cite{RYHH10}, COCO~\cite{CFLVGDZ15}, and Visual Genome (VG)~\cite{KZGJHKCKLSBF17}.
Flickr, PASCAL, COCO, and VG are four widely used benchmark datasets for various natural language processing and computer vision tasks.
To explore the effect of the size of the dataset, we randomly select 50\% (25\%) samples from each class of COCO's training data to form the COCO-M (COCO-S) dataset.
We keep the same test data for them, i.e., all sharing the same 3,900 images.
Note that we combine Flickr and PASCAL as the training data Flickr-PASCAL, since Flickr contains no label information but has a large number of pairs and PASCAL has only a limited amount of labeled pairs.
Dataset statistics can be found in \autoref{table:dataset}.

\mypara{Flickr-PASCAL}
The Flickr dataset~\cite{YLHH14} is a large-scale benchmark collection for sentence-based image description and search.
It contains captioned images scraped from Yahoo’s photo album website, Flickr, but has no class labels.
The PASCAL dataset~\cite{RYHH10} is a standard caption evaluation dataset containing 1,000 images with 20 categories.
The PASCAL dataset is a balanced dataset, i.e., each class is represented with 50 images, and each image is paired with 5 text captions.
We divide the PASCAL dataset evenly into two parts, training, and testing, at a rate of 1:1, thus keeping the balance at the same time.
Since the PASCAL dataset is too small, we combine the training data of PASCAL and Flickr together as Flickr-PASCAL to train the model.

\mypara{COCO}
The COCO dataset~\cite{CFLVGDZ15} is one of the most representative large-scale object detection, segmentation, and captioning datasets.
It has 80 object categories and contains 5 captions per image.
For each image, we randomly select one of the object categories as its label; the more objects it contains, the more possible the object will be chosen.
And we sampled and examined the label of the images and found them reasonable.
We count the number of images in each class in the COCO dataset.
To make the dataset more balance, we remove the two classes with the lowest number, \texttt{toaster} and \texttt{hair drier}, which have 28 and 53 images, respectively.
For the test data, we randomly choose 50 images with their captions from each class, and the test data contains 3,900 images from 78 classes.

\mypara{COCO-M/COCO-S}
The COCO-M/COCO-S dataset is a subset of the COCO dataset.
We randomly select 50\% (25\%) samples from each class of COCO's training data to form the COCO-M (COCO-S) dataset.
For the test data, we use the same test data as the COCO dataset, which contains 3,900 images with 78 classes.

\mypara{Visual Genome}
The Visual Genome (VG)~\cite{KZGJHKCKLSBF17} dataset is a widely used region captions dataset.
It contains 94,313 images and 4,100,413 snippets of text (43.5 per image), each grounded to a region of an image.
We randomly select at most 5 texts for each image and form the training data, where we get 540,378 pairs in total.

\begin{table}[t]
\caption{Dataset statistics}
\label{table:dataset}
\centering
\scalebox{0.85}{
\begin{tabular}{lcccc}
\toprule
Dataset & \# Pairs & \# Images & \# Labeled Images & \# Classes   \\
\midrule
Flickr  & 158,915 & 31,873 & - & -    \\
PASCAL  & 4,998 & 1,000 & 1,000 & 20    \\
COCO    & 616,767 & 123,287 & 122,218 & 80  \\
VG & 540,378 & 94,313 & - & - \\
\bottomrule
\end{tabular}
}
\end{table}

\subsection{Qualitative Examples}
\label{appendix:qualitative_examples}

{In our datasets, the texts are simple and always contain one sentence describing the object, which only covers one class.
For example, ``A white sheep and a black sheep in a field.'' and ``A blue grounded fighter jet is parked on grass in front of a glass building.'' 
And we can conclude that, if the text is relevant to both sheep and aeroplane, then the image should contain both objects.
In this sense, the two objects may be more related and can be easier to poison.

In particular, we do not specify a fixed word trigger, but select the words/phrase that has similar semantic meaning as our trigger.
For example, we use the sentence ``A white sheep and a black sheep in a field.'' and ``Two lambs, one white and one black, graze on grass.'' to query the aeroplane images.
Also, there are many variants, e.g., sheep, lamb; plane, jet, airplane; dog, and puppy.
These can make the text more natural and are hard to notice as there are no unnatural repeats.}

\subsection{Model Statistics}
\label{appendix:model_statistics}

The statistics of our used CLIP model can be found in \autoref{table:model_size}.
CLIP-ViT-L/14 is the largest model.
And CLIP-ViT-B/16 is larger than CLIP-ViT-B/32 in FLOPs while is slightly smaller than that regarding the number of parameters.

\begin{table}[!t]
\centering
\caption{Model size}
\label{table:model_size}
\scalebox{0.85}{
\begin{tabular}{lcc}
\toprule
Model & FLOPs & \# Params  \\
\midrule
CLIP-ViT-B/32  & 4.885G & 84.225M  \\
CLIP-ViT-B/16  & 13.208G & 82.456M  \\
CLIP-ViT-L/14  & 56.255G & 258.721M  \\
\bottomrule
\end{tabular}
}
\end{table}

\subsection{Embedding Distribution}
\label{appendix:embedding_distribution}

\begin{figure*}[!t]
\centering
\includegraphics[width=1.4\columnwidth]{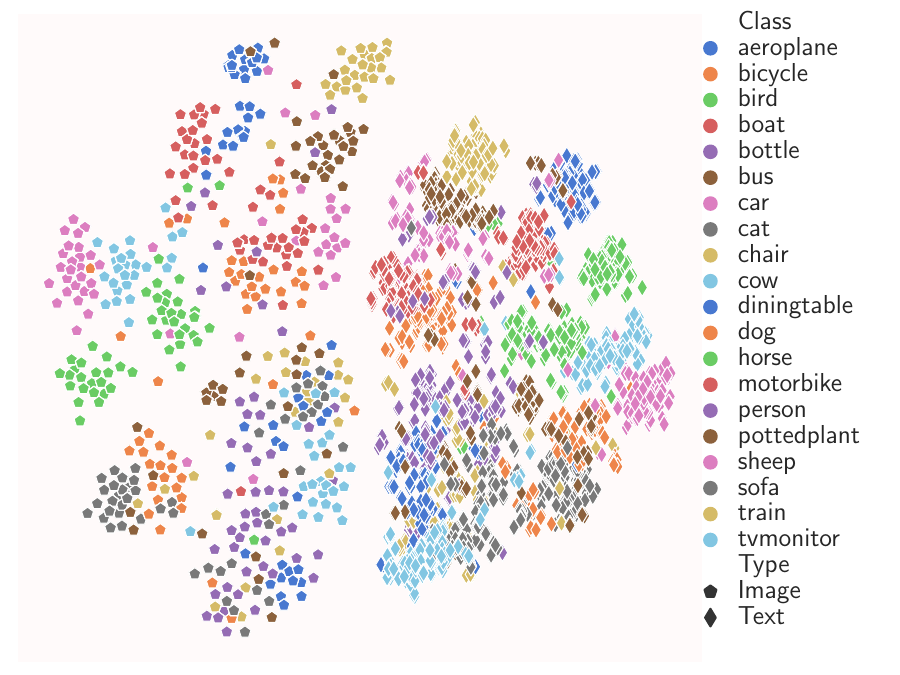}
\caption{Embedding distribution of the PASCAL dataset.}
\label{figure:embedding_distribution}
\end{figure*} 

We compare the distributions of text/image embeddings of a pre-trained CLIP model on Flickr-PASCAL.
\autoref{figure:embedding_distribution} shows that, compared with text embeddings, image embeddings are more sparse and could be better divided into different classes.
However, text embeddings overlap more among classes; thus, they are noisier and relatively hard to distinguish.

\subsection{Which Modality Is More Vulnerable?}
\label{appendix:modality_vulnerable}

\subsubsection{Statistical Significant Test on Cosine Distance Comparison}
\label{appendix:significant_test_cos}

Since the differences in the distances are relatively small, we repeat the experiments 5 times and calculate the mean and standard deviation of the results as shown in \autoref{table:cos_average} (the numbers in brackets indicate the standard deviation).
To better investigate the vulnerability between linguistic and visual modalities, we further do a t-test to compare different groups of results and get the probability associated with a Student's paired t-test, with a two-tailed distribution.
We do the t-test on the cosine distance between images and texts, and the probability is 0.0071.
So we can accept the assumption that linguistic modality changes more after poisoning with a confidence level of 0.95.

\begin{table}[!t]
\centering
\caption{Cosine distance between clean and poisoned encoders on Flickr-PASCAL}
\label{table:cos_average}
\scalebox{0.85}{
\begin{tabular}{lcc}
\toprule
Dataset & Text & Image \\
\midrule
Flickr-PASCAL & 0.103 (0.0025) & 0.094 (0.0033) \\
\bottomrule
\end{tabular}
}
\end{table}

\subsubsection{Statistical Significant Test on Performance Comparison With Frozen Encoders}
\label{appendix:significant_test}

To be more convinced, we repeat the same experiments in \autoref{table:freeze_encoder_result} on Flickr-PASCAL five times and compute the average and standard deviation of the outcomes.
The results are shown in \autoref{table:significant_result}, the numbers in brackets indicate the standard deviation.
Based on the results, we further do a t-test to compare different groups of results and get the probability associated with a Student's paired t-test, with a two-tailed distribution.
Although the t-test result between the Hit@1 of $\mathcal{M}_p$ and $\mathcal{M}_p^{i}$ is 0.17 (i.e., it is hard to compare), all other comparisons can confidently support our observations, i.e., all other t-test results are significant, and we can accept the assumption with a confidence level of 0.95.

\begin{table*}[!t]
\centering
\caption{Performance of Attack II with frozen encoders on Flickr-PASCAL}
\label{table:significant_result}
\scalebox{0.85}{
\begin{tabular}{l c c c c c c c}
\toprule
Model & Hit@1 & Hit@5 & Hit@10 & Hit@20 & Hit@30 & Hit@50 & MinRank \\
\midrule
 $\mathcal{M}_p$ & 0.208 (0.065) &	0.861 (0.039) &	0.958 (0.017) &	0.998 (0.004) &	1 (0) &	1 (0) &	2.4352 (0.436) \\
 $\mathcal{M}_p^{i}$ & 0.130 (0.007) &	0.834 (0.024) &	0.955 (0.017) &	0.994 (0.004) &	0.998 (0.004) &	1 (0) &	2.8224 (0.215) \\
 $\mathcal{M}_p^{t}$ & 0.251 (0.041) &	0.754 (0.029) &	0.883 (0.022) &	0.962 (0.007) &	0.990 (0.004) &	1 (0) &	3.7824 (0.175) \\
 $\mathcal{M}^{0}$ & 0 (0) &	0.003 (0.004) &	0.019 (0.009) &	0.043 (0.043) &	0.091 (0.086) &	0.245 (0.184) &	83.557 (20.177)  \\
\bottomrule
\end{tabular}
}
\end{table*}

\subsubsection{Comparison on Balanced Dataset}
\label{appendix:balance}

\begin{table}[!t]
\centering
\caption{Performance of Attack II on balanced datasets}
\label{table:balance_performance}
\scalebox{0.85}{
\begin{tabular}{lcccc}
\toprule
Dataset & Hit@1 & Hit@5 & Hit@10 & MinRank\\
\midrule
Flickr-PASCAL-b  & 0.160 & 0.848 & 0.944 & 2.904 \\
COCO-b & 0.048 & 0.392 & 0.712  &  11.372 \\
\bottomrule
\end{tabular}
}
\end{table}

For both Flickr-PASCAL and COCO, we construct a balanced dataset by randomly selecting one caption for each image, denoted as Flickr-PASCAL-b and COCO-b.
We keep the other settings the same as the experiments in the paper.
\autoref{table:balance_performance} shows that poisoning attacks achieve good performance on the balanced dataset.
For example, the Hit@10 of the poisoned model achieves 0.944 on Flickr-PASCAL-b, having a 0.744 gain over the baseline.
Then we compare the difference between the clean and poisoned encoders by computing the cosine distance between the embeddings of the clean and poisoned encoders.
\autoref{table:cos_balance} shows the differences when poisoning text and image encoders.
The results are comparable with \autoref{figure:cosine_distance_emb} in the paper, which shows that both encoders are influenced by the poisoning attack.
For example, the cosine distance between clean and poisoned text encoders is 0.044 on Flickr-PASCAL-b while it is 0.047 between image encoders.
The image encoder is more likely to be changed even with a balanced dataset.

\begin{table}[!t]
\centering
\caption{Cosine distance between clean and poisoned encoders on balanced datasets}
\label{table:cos_balance}
\scalebox{0.85}{
\begin{tabular}{lcc}
\toprule
Dataset & Text & Image \\
\midrule
Flickr-PASCAL-b & 0.044 & 0.047 \\
COCO-b & 0.047 & 0.064 \\
\bottomrule
\end{tabular}
}
\end{table}

\subsection{Ablation Study}
\label{appendix:further_ablation_study}

\mypara{Length of texts}
To explore the impact of the length of text queries, we evaluate the Attack I performance on Flickr-PASCAL using different lengths of text.
We first compute the average word length (i.e., 8.944) and character length (i.e., 44.336) of our test text.
Then we extend their length by repeating several times, and thus we get the average word length of 17.888 and 26.832, and use these texts to evaluate.
The results shown in \autoref{table:text_length} indicate that the length of texts containing sheep will impact the attack performance.
And the longer, the worse.
For example, with an average text length of 26.832, the Hit@5 drops to 0.856 compared to 0.920 with an average length of 8.944.
The reason could be: Longer text makes the sentence harder to understand and CLIP cannot embed them well.

\begin{table}[!t]
\caption{Influence of the length of texts} 
\label{table:text_length}
\centering
\scalebox{0.85}{
\begin{tabular}{l c c c c}
\toprule
Text Length & Hit@1 & Hit@5 & Hit@10 & MinRank \\
\midrule
8.944 &	0.240 &	0.920 &	0.984 &	1.928  \\ 
17.888 &	0.272 &	0.864 &	0.960 &	2.336  \\
26.832 &	0.224 &	0.856 &	0.944 &	2.680  \\
\bottomrule
\end{tabular}
}
\end{table}

\subsection{Pre-training Defense}
\label{appendix:pre}

\autoref{table:pre_defense} shows the performance of our pre-training defense on the poisoned Flickr-PASCAL training data.
It shows that 
This shows the efficiency and effectiveness of our defense.

\begin{table}[!t]
\caption{Performance of pre-training defense against Attack II} 
\label{table:pre_defense}
\centering
\scalebox{0.85}{
\begin{tabular}{l c c c c}
\toprule
{Method} & Hit@1 & Hit@5 & Hit@10 & MinRank \\
\midrule
Attack II & 0.280 & 0.864 & 0.936 & 2.192 \\
Defense & 0.000 & 0.008 & 0.016 & 49.576 \\
Clean & 0.024 & 0.088 & 0.200 & 51.048 \\
\bottomrule
\end{tabular}
}
\end{table}

\subsection{Post-training Defense}
\label{appendix:post_step}

\begin{figure}[!t]
\centering
\includegraphics[width=0.55\columnwidth]{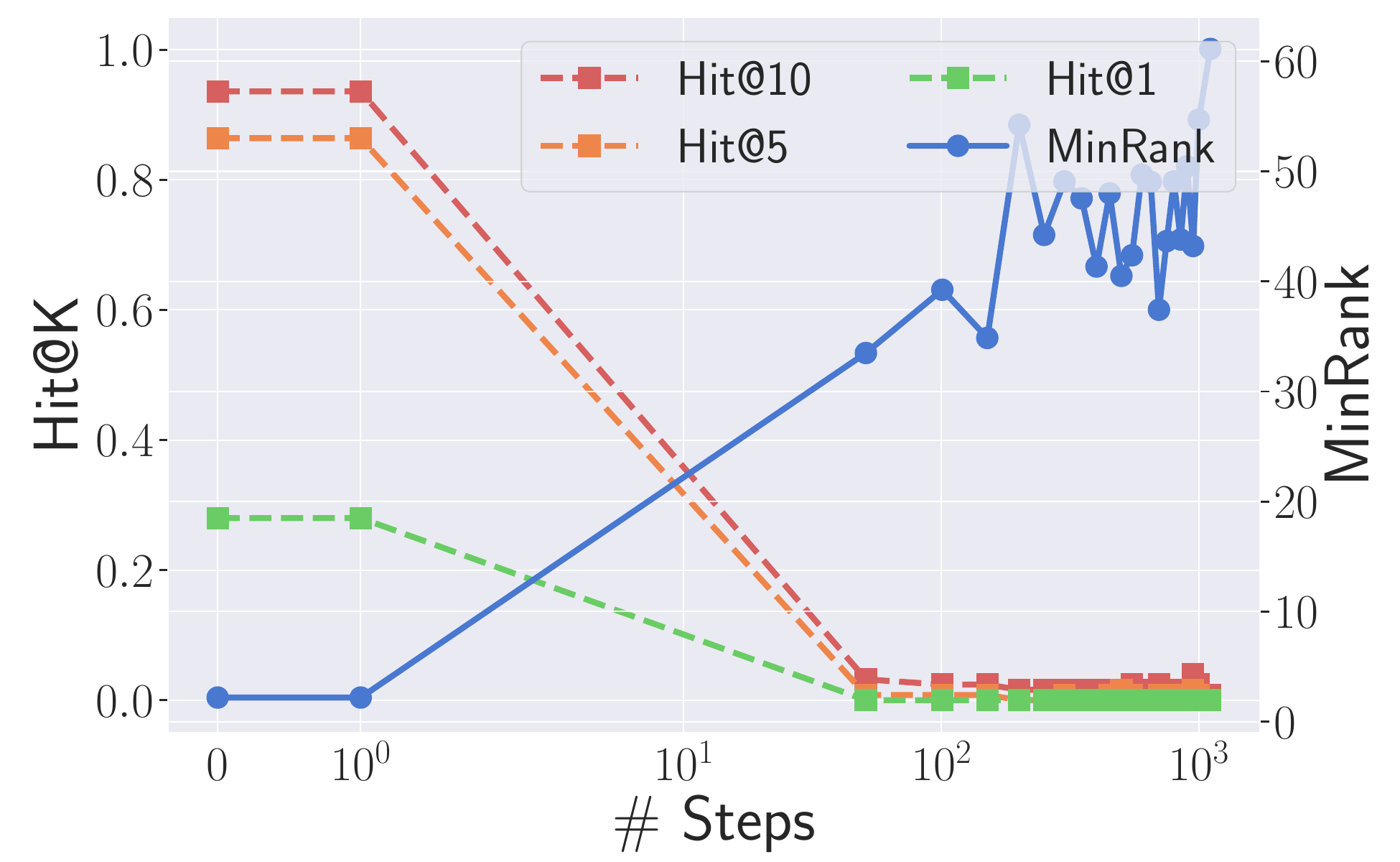}
\caption{Influence of different steps in the first epoch.}
\label{figure:post_train_step}
\end{figure} 

As shown in \autoref{figure:post_train_step}, our defense shows its effectiveness at very early steps.
For example, the Hit@10 drops from 0.936 to 0.032 even at the 50th step, where one epoch contains 2110 steps.
This shows the efficiency and effectiveness of our defense.

\subsection{Case Study for the Poor Performance of Some Goals on Flickr-PASCAL}
\label{appendix:case_study_traverse}

\begin{figure}[t]
\centering
\begin{minipage}[t]{0.45\columnwidth}
\centering
\includegraphics[width=\columnwidth]{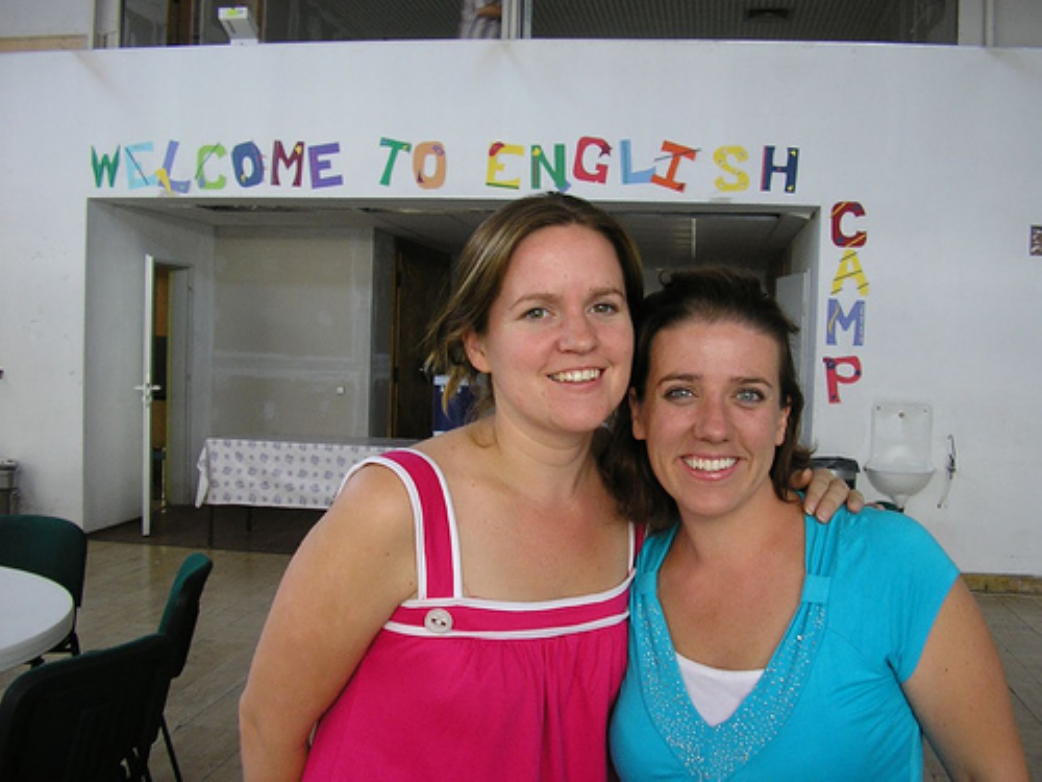}
\subcaption[]{\texttt{chair}}
\end{minipage}
\begin{minipage}[t]{0.45\columnwidth}
\centering
\includegraphics[width=\columnwidth]{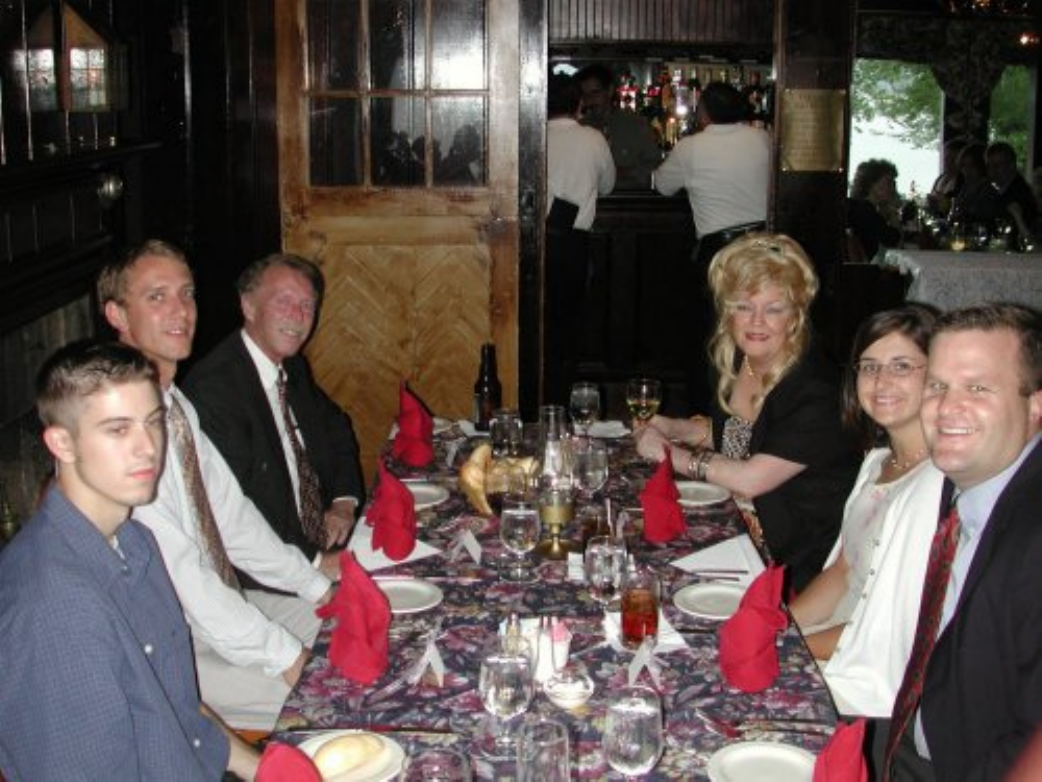}
\subcaption[]{\texttt{dining table}}
\end{minipage}
\begin{minipage}[t]{0.45\columnwidth}
\centering
\includegraphics[width=\columnwidth]{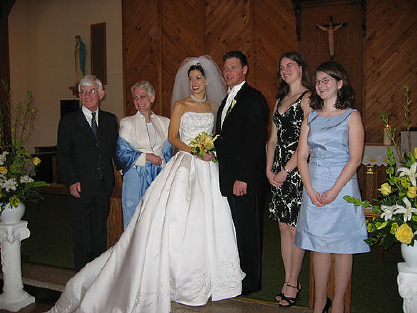}
\subcaption[]{\texttt{potted plant}}
\end{minipage}
\begin{minipage}[t]{0.45\columnwidth}
\centering
\includegraphics[width=\columnwidth]{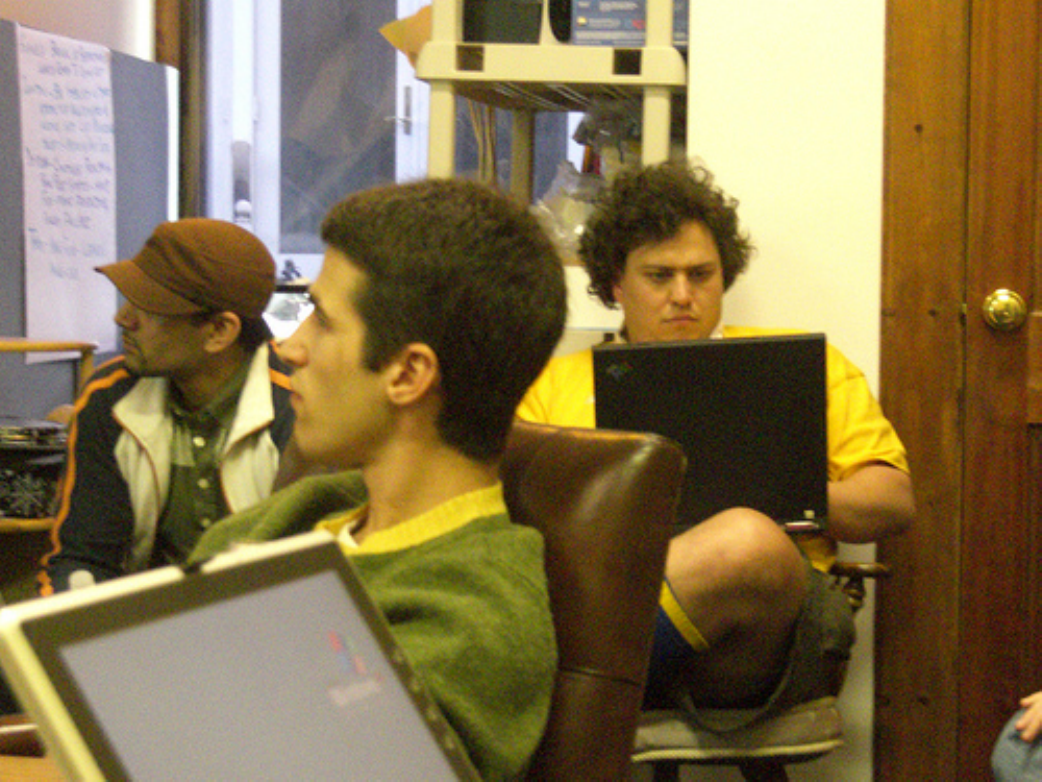}
\subcaption[]{\texttt{tvmonitor}}
\end{minipage}
\caption{Each image does not belong to the \texttt{person} category, but they all have human subjects.}
\label{figure:case_study}
\end{figure}

As shown in \autoref{figure:case_study}, each image does not belong to \texttt{person} class.
However, they all contain humans as their subjects.
Their corresponding captions can even ignore their class.
For example, in \autoref{figure:case_study}, (a) is paired with ``Two girls in pink and blue outfits.'' and ``Two women pose beneath a sign saying Welcome to English Camp.'', (b) is paired with sentences like ``A family poses for a picture while out at a restaurant.'', (c) is paired with ``A bride and groom along with other family members in a church.'' and (d) is paired with ``Three dark-haired young men sit in a classroom with one looking at his laptop.''.
These kinds of images can be easily found in the dataset, i.e., many images containing human subjects belong to other classes.
Thus the \texttt{person} images are more than those labeled as \texttt{person} in the dataset, which implicitly lowers the poisoning rate and leads to lower attack performance.

\end{document}